\documentclass[11pt,twoside,hyphens]{article} 
\usepackage{asp2010}

\resetcounters

\begin{document}

\title{CANFAR+Skytree: A Cloud Computing and Data Mining System for Astronomy}
\author{Nicholas M. Ball$^1$
\affil{$^1$National Research Council Canada, 5071 West Saanich Road, Victoria, BC V9E 2E7}}

\begin{abstract}
This is a companion Focus Demonstration article to the CANFAR+Skytree poster (\citeauthor{ball:adass12poster} \citeyear{ball:adass12poster}, this volume), demonstrating the usage of the Skytree machine learning software on the Canadian Advanced Network for Astronomical Research (CANFAR) cloud computing system. CANFAR+Skytree is the world's first cloud computing system for data mining in astronomy.
\end{abstract}

\section{Introduction}

CANFAR\footnote{\url{http://canfar.phys.uvic.ca}} \citep{gaudet:canfar} is the cloud computing system of the Canadian Astronomy Data Centre (CADC). It is the first system designed to provide this capability to astronomers. Skytree is the world's most advanced machine learning software. It acts as a machine learning server to allow advanced data mining on large data. The CANFAR+Skytree combination allows Skytree to be run on up to 500 cores simultaneously, the current size of the CANFAR system. In this paper, we reproduce the Focus Demonstration session given at the conference, showing an example Skytree run, how to access and use CANFAR, and how to use the two in concert.

\section{Running Skytree}

Skytree can be run interactively on a UNIX or Mac OS X system. Installation is by unzipping the tarfile into a directory

\begin{verbatim}
tar -zxf SkytreeServer11.3.2.tgz
\end{verbatim}

This results in a directory containing the skytree-server executable, an example dataset, a .lic license file, and some others.

\subsection{Example: Nearest Neighbors}

We show the running of the nearest neighbors algorithm, {\tt allkn} on the example dataset supplied with the software. The dataset is from the Sloan Digital Sky Survey, reflecting the company's academic roots and links to astronomy. It contains just under 100,000 rows (galaxies), and four SDSS colors.

Begin by selecting suitable rows:

\begin{verbatim}
cd SkytreeServer11.3.2/datasets
awk -F, '(NF==3 || NF==7)' sdss100kx4.skytree \
   > sdss100kx4.subsample.skytree
cd ..
\end{verbatim}

This is typical of analysis using Skytree: as with any data mining, one prepares the data first. Some data preparation and results analysis tools are now available with the software, but the machine learning invocation remains separate, on whatever file it is passed. Input files are typically ASCII format. The .skytree represents an explicit header style in which datatypes are given that enables some algorithms to run faster, but the file is otherwise ASCII CSV.

\begin{figure}
\plotone{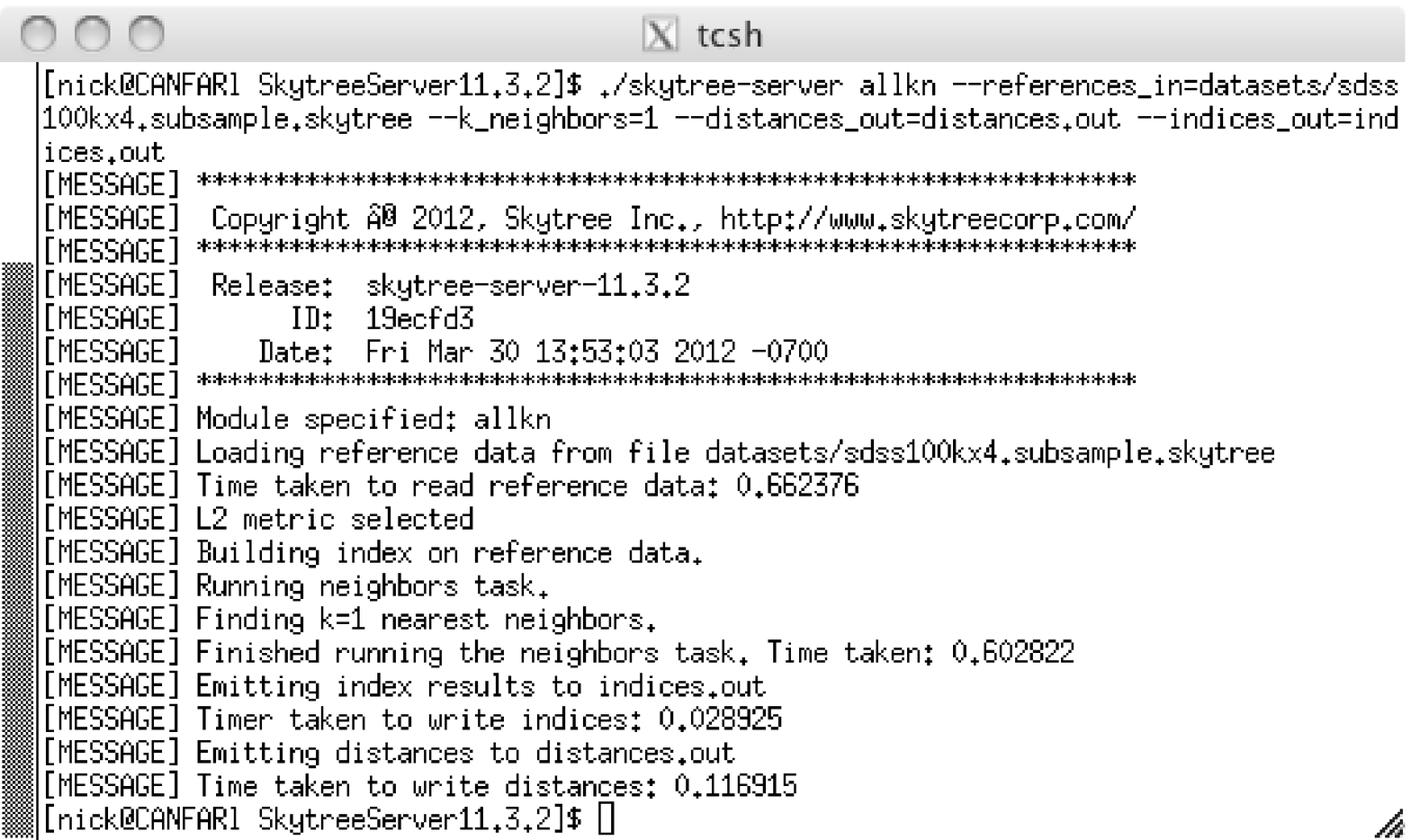}
\caption{Typical Skytree invocation on the terminal, showing the {\tt allkn} run detailed in the text. \label{Fig: Terminal}}
\end{figure}

We then run {\tt allkn}:

\begin{verbatim}
./skytree-server allkn \
   --references_in=datasets/sdss100kx4.subsample.skytree \
   --k_neighbors=1 \
   --distances_out=distances.out \
   --indices_out=indices.out
\end{verbatim}

The program is invoked via the {\tt skytree-server} executable, the algorithm name (in this case {\tt allkn}), and passed arguments as appropriate to the algorithm. In this case, the input file, {\tt references\_in}, the number of neighbors to find, 1, and the neighbor distances and file positions as output. The typical appearance of this run in the terminal is shown in Figure \ref{Fig: Terminal}. Each algorithm is fully documented if invoked with the {\tt --help} argument.

Once run, we have obtained the neighbor distances, and, via the indices, which objects are the neighbors. These may be cast into a suitable form for visualization, e.g.

\begin{verbatim}
paste -d \\0 distances.out indices.out \
   datasets/sdss100kx4.subsample.skytree > tmp.csv
sed s/'header,double:1,header,unsigned_int:1,header,meta:3,\
   double:4'/'#nn,ind,label,target,id,ug,gr,ri,iz'/g tmp.csv \
   > results_allkn.csv
\end{verbatim}

\noindent which may be visualized in a program such as TOPCAT \citep{taylor:topcat}. Again, this is typical of an analysis with Skytree: it outputs results, which are then further processed.

In this case, if one histograms the distances, selects those with large distances, and plots a color-color plot, e.g., $u-g$ versus $g-r$ ($ug$ vs. $gr$), it is clear that {\tt allkn} has found outliers. Obviously such a measure in isolation is crude (one might want to calculate, for example, the local outlier factor), but it exemplifies the kind of analysis that can be rapidly built up using data mining.

\section{Running Software on CANFAR (Including Skytree)}

To access CANFAR requires a CADC account, and a CANFAR account. These are set up via the CADC webpage at \url{http://www.cadc-ccda.hia-iha.nrc-cnrc.gc.ca}, and by request to {\tt CanfarHelp@nrc-cnrc.gc.ca} . Once given an account, access CANFAR via ssh:

\begin{verbatim}
desktop> ssh <username>@canfar.dao.nrc.ca
\end{verbatim}

This places the user on the CANFAR head node, from which it is possible to utilize software interactively, and run short processing jobs (e.g., a half hour or less). X Windows and X-forwarding is supported. Thus, one may install software as desired, including Skytree, and run it as above. Detailed usage of CANFAR is documented on the wiki, at \url{http://canfar.phys.uvic.ca/wiki}.

\subsection{Virtual Machines}

Rather than installing software in one's home directory on the CANFAR head node, the bulk of the interaction with the system is via a virtual machine (VM). The VM is created by the user, who then has full root access to it. Access is via ssh:

\begin{verbatim}
CANFAR> vmcreate <vmname>
CANFAR> vmssh <vmname>
\end{verbatim}

To shut down the VM, use {\tt vmstop}. One a VM exists, one does not {\tt vmcreate} it again, but starts it using {\tt vmstart}.

\subsection{VOSpace}

CANFAR has implemented, via International Virtual Observatory Alliance protocols, a filesystem, VOSpace, that gives CANFAR users access to hundreds of terabytes of persistent storage. Access to VOSpace requires an X.509 certificate, which can be obtained by the user via \url{http://www.cadc-ccda.hia-iha.nrc-cnrc.gc.ca/cadcbin/auth/archive/accountDetails.pl}. VOSpace can be mounted as a filesystem, which enables it to be treated as another directory tree, and accessed from one's desktop, the CANFAR head node, one's VM, or a batch job.

\subsection{Batch Jobs}

Batch jobs are managed by the Condor scheduling system. To prepare a batch job, a Condor submission file and a calling script are created on the CANFAR head node, which in turn calls a script on the VM. To submit a job, the VM is shutdown, and the Condor submission command is given:

\begin{verbatim}
CANFAR> vmstop <vmname>
CANFAR> condor_submit <job>.condor
\end{verbatim}

One may then monitor the execution of the job via the usual Condor commands, e.g., {\tt condor\_q <username>}. Jobs have access to scratch disk staging space, and the results are copied back to VOSpace.

\subsection{Running CANFAR+Skytree}

Skytree is invoked on the command line or a script as part of one's analysis. Running in batch allows up to 500 instances of Skytree to be run simultaneously.

\section{Conclusions}

CANFAR+Skytree represents world's first cloud computing system for data mining in astronomy, and is open for use by any interested member of the astronomical community. For further details on usage, see the poster paper (\citeauthor{ball:adass12poster} \citeyear{ball:adass12poster}, this volume), or visit the CANFAR+Skytree website at \url{https://sites.google.com/site/\-nickballastronomer}.

\acknowledgements This research used the facilities of the Canadian Astronomy Data Centre, operated by the National Research Council of Canada with the support of the Canadian Space Agency. Funding for CANFAR was provided by CANARIE via the Network Enabled Platforms Supporting Virtual Organisations program. The author thanks D. Schade, A. Gray and M. Hack for their contributions to this work.

\bibliography{F5_refs}

\end{document}